\documentclass{svjour3}                     % onecolumn (standard format)
\smartqed  % flush right qed marks, e.g. at end of proof
\usepackage{graphicx}
\begin{document}

\title{Evanescent wave transport and shot noise in graphene: ballistic regime and effect of disorder \thanks{This work was supported by the Academy of Finland, the EU CARDEQ contract FP6-IST-021285-2 and the NANOSYSTEMS contract with the Nokia Research Center.}
}


\author{R. Danneau \and F. Wu \and M.F. Craciun \and S. Russo \and M.Y. Tomi \and J. Salmilehto \and A.F. Morpurgo \and P.J. Hakonen}


\institute{R. Danneau . F. Wu . M.Y. Tomi . J. Salmilehto . P.J. Hakonen
           \at
              Low Temperature Laboratory, Helsinki University of Technology, Espoo, Finland \\
              Tel.: +358-5-03442423\\
              Fax: +358-9-4512969\\
              \email{r.danneau@boojum.hut.fi}           %  \\
           \and
           M.F. Craciun . S. Russo . A.F. Morpurgo \at
              Kavli Institute of Nanoscience, Delft University of Technology, Delft, The Netherlands}

\date{Received: date / Accepted: date}

\maketitle
%We show that in graphene $1/f$ noise is as high as in carbon nanotubes. More astonishing, we found that $1/f$ noise is independent on the charge carrier density in short and wide graphene strip
\begin{abstract}
We have investigated electrical transport and shot noise in graphene field effect devices. In large width over length ratio $W/L$ graphene strips, we have measured shot noise at low frequency ($f$ = 600--850 MHz) in the temperature range of 4.2--30 K. We observe a minimum conductivity of $\frac{4e^{2}}{\pi h}$ and a finite and gate dependent Fano factor reaching the universal value of 1/3 at the Dirac point, i.e. where the density of states vanishes. These findings are in good agreement with the theory describing that transport at the Dirac point should occur via evanescent waves in perfect graphene samples with large $W/L$. Moreover, we show and discuss how disorder and non-parallel leads affect both conductivity and shot noise.
\end{abstract}

\section{Introduction}
\label{intro}

Graphene is a unique two-dimensional material. Its recent discovery has spawned great interest in the scientific community \cite{geim2007}. Graphene is a gapless semiconductor: the conduction and valence bands touch in two inequivalent points (K and K', usually called Dirac points) where the density of states vanishes. However, the conductivity at the Dirac point remains finite. Indeed, at the Dirac point, it has been theoretically shown that in perfect graphene, the conduction occurs only via evanescent waves, i.e. via tunneling between the leads \cite{katsnelson2006a,tworzydlo2006}. We present measurements of conductivity and noise in graphene strips that support this theory.

Effect of disorder, interactions or carrier statistics can be assessed accurately by probing shot noise in mesoscopic devices \cite{blanter2000}. These out of equilibrium current fluctuations arise from the granular nature of electron charges. Indeed, shot noise provides a powerful tool to reveal information on fundamental conduction properties of low-dimensional systems which are not accessible via conventional dc transport measurements. For example, such current fluctuations have been used to show that fractional charges can carry current \cite{saminadayar1997,depicciotto1997}, to demonstrate the fermionic nature of electrons \cite{henny1999,oliver1999}, and to study many-body phenomena in mesoscopic physics \cite{safonov2003,roche2004,dicarlo2006}.

%First, we show that the $1/f$ noise in such graphene strips is as high as in single walled carbon nanotubes.

In this article, we address a study of the noise in short and wide graphene strips. Using a home-made low-noise amplification set-up, we measure shot noise as a function of gate voltage in two-terminal field-effect graphene devices. We show that the transport via evanescent wave theory is in good agreement with our results on large width over length ratio $W/L$ samples when the distance between the leads is 200 nm. We show how the disorder affects the conductivity and the shot noise. Additionally, we have measured shot noise in the case of non-rectangular systems, i.e. when the leads are not parallel.

\section{Transport via evanescent waves at the Dirac point}

Using the scattering matrix formalism (see \cite{dattabook}), one can express the carrier transport of a mesoscopic system. The conductance of each quantum channel carrying current can be written as $G = g\frac{e^2}{h}T$, where $g$ is the degeneracy (spin and valley) of the system and $T$ the electron transmission probability. When the system is biased, current fluctuations appear and for a single channel they can be described by $\langle (\delta I)^2 \rangle = 2e\langle I \rangle(1-T)$. Shot noise is due to the discreteness of charge \cite{blanter2000}. It can only be detected when the electron-phonon inelastic scattering length $L_{e-ph}$ and electron-electron inelastic scattering length $L_{e-e}$ are much larger than any sample dimension \cite{beenakker1992,nagaev1992,shimuzu1992,steinbach1996}. Then, one can write the noise power spectrum which is proportional to the product of the transmission $T$ and the reflection $R = 1 - T$, summed over the $N$ channels:
\begin{eqnarray}
S_{I} = \frac{2e^3|V|}{h}\sum_{n=0}^{N-1} T_n(1-T_n)
\end{eqnarray}
In the limit of low transparency $T_n \ll 1$,
\begin{eqnarray}
S_{I} = S_{P} = \frac{2e^3|V|}{h}\sum_{n=0}^{N-1} T_n = 2e\langle I \rangle
\end{eqnarray}
defining a Poissonian noise induced by independent and random electrons like in tunnel junctions \cite{blanter2000}. The common way to quantify shot noise is to use the Fano factor $\mathcal{F}$ which is the ratio between the measured shot noise and the Poissonian noise:
\begin{eqnarray}
 \mathcal{F} = \frac{S_I}{S_P} = \frac{S_I}{2e\langle I \rangle} = \frac{\sum_{n=0}^{N-1}T_n(1-T_n)}{\sum_{n=0}^{N-1} T_n}
\end{eqnarray}
Then, for a Poissonian process $\mathcal{F} = 1$ (at small transparency $T_n \rightarrow 0$, i.e. when transport occurs via electron tunneling) while $\mathcal{F} = 0$ in the ballistic regime (i.e. in the perfect transmission case $T_n \rightarrow 1$) and $\mathcal{F} = 1/3$ in the case of a diffusive system.

\begin{figure}[h]
\includegraphics[width=18pc]{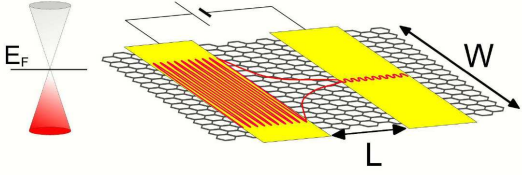}\hspace{2pc}%
\begin{minipage}[b]{9pc}\caption{\label{evanescent}Schematics representing transport via evanescent waves in perfect graphene with large aspect ratio $W/L$. Evanescent states transport occurs when the Fermi energy is set at the Dirac point. Away from the Dirac point, transport occurs via propagating plane waves.}
\end{minipage}
\end{figure}

In graphene, it has been demonstrated that transport at the Dirac point may occur via electronic evanescent waves \cite{katsnelson2006a,tworzydlo2006} (illustration in Figure \ref{evanescent}). Tworzyd{\l}o \emph{et al.} used heavily-doped graphene leads and the wave function matching method to directly solve the Dirac equation in perfect graphene with length $L$ and width $W$ \cite{tworzydlo2006}. They found that for armchair edges, the quantization condition of the transverse wave vector is defined by $k_{y,n}=\frac{(n+\alpha)}{W}\pi$ where $\alpha = 0$ or $\frac{1}{3}$ for metallic or semiconducting armchair edges. At the Dirac point, the transmission coefficient reads:
\begin{eqnarray}
T_n^{Dirac} = \frac{1}{\cosh^{2}(\pi (n+\alpha)\frac{L}{W})}
\end{eqnarray}
As we can see, at the Dirac point, graphene has a similar bimodal distribution of transmission eigenvalues as in diffusive systems \cite{blanter2000}. Finally, in the case of large $W/L \rightarrow \infty$, the mode spacing becoming small, one can replace the sum over the $N $ channels by an integral over the transverse wave vector component $k_y$ to obtain the conductivity and the Fano factor for metallic armchair edges systems:
\begin{eqnarray}
\sigma_{Dirac} = G_{Dirac}\frac{L}{W} = \frac{4e^2}{h} \frac{L}{W}\int_{0}^{\infty}\frac{dk_y}{\cosh^{2}(k_yL)}=\frac{4e^2}{\pi h}
\end{eqnarray}
\begin{eqnarray}
\mathcal{F}_{Dirac} = \frac{\sum_{n=0}^{N-1}T_n^{Dirac}(1-T_n^{Dirac})}{\sum_{n=0}^{N-1} T_n^{Dirac}}\equiv\frac{\int_{0}^{\infty}\frac{dk_y}{\cosh^{2}(k_yL)}(1-\frac{1}{\cosh^{2}(k_yL)})}{\int_{0}^{\infty}\frac{dk_y}{\cosh^{2}(k_yL)}}=\frac{1}{3}
\end{eqnarray}
At the Dirac point, in the case of coherent carrier transport, the conductivity is minimum ($\sigma_{Dirac} = \frac{4e^2}{\pi h}$) and the Fano factor is maximum ($\mathcal{F}_{Dirac} = \frac{1}{3}$) \cite{tworzydlo2006}. It is important to note that metallic leads \cite{schomerus2007} do not affect the evanescent wave theory. However, both conductivity and Fano factor are no longer respectively minimum and maximum when the transport in incoherent \cite{sonin2008}. Moreover, in large samples the minimum conductivity has been measured around $\sigma_{Dirac} = \frac{4e^2}{h}$ \cite{geim2007,novoselov2005,tan2007}, which could be explained by the presence of disorder \cite{badarson2007}. By tuning the carrier density, the Fermi level is moved away from the Dirac point where the density of states is no longer zero. At large density the number of conducting channels increases, the evanescent states are then accompanied by propagating states, the conductivity rises while the Fano factor decreases \cite{tworzydlo2006}. Note that the duality between evanescent and propagating waves could be studied using multi-probes and cross-correlation measurements \cite{laakso2008}. The Fano factor for a bilayer system has been predicted to be $\frac{1}{3}$ as well \cite{snyman2007} or $1-\frac{2}{\pi}$ \cite{katsnelson2006}, i.e. very close to $\frac{1}{3}$. We note that in graphene \emph{pn}-junctions the Fano factor is also very close to $\frac{1}{3}$ \cite{cheianov2006} and takes values depending on the Landau level filling factors under magnetic field \cite{abanin2007}. Finally, transport at the Dirac point remains not fully understood. The fact that the distribution function of the transmission probability of the evanescent states is exactly the same as the propagated states in diffusive systems is still unexplained. This resulting exotic shot noise for a ballistic system might be related to relativistic quantum dynamics of confined Dirac fermions which are known to exhibit a jittering motion called \emph{zitterbewegung} \cite{katsnelson2006a,tworzydlo2006}.

\begin{figure}[htbp]
\begin{center}\scalebox{0.25}{\includegraphics{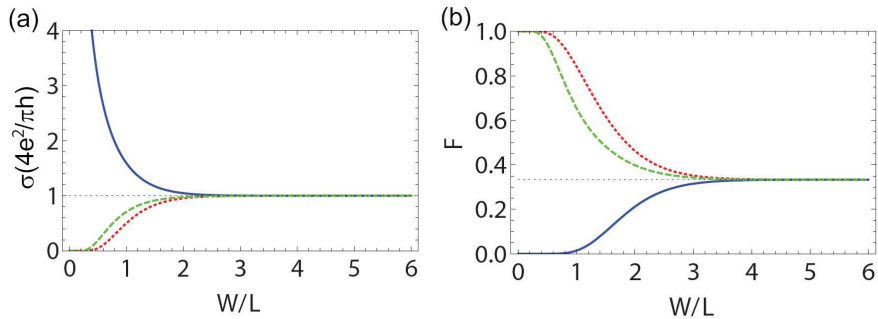}}\end{center}
{\caption{\label{evanescent-sigma-F}(a) Minimum conductivity and (b) Fano factor at the Dirac point versus the width over length ratio $W/L$ calculated using the evanescent wave theory \cite{tworzydlo2006}, for three different boundary conditions: for metallic armchair edges in blue solid line ($\alpha = 0$), for semiconducting armchair edges in green dashed line ($\alpha = \frac{1}{3}$) and smooth edges in red dotted line ($\alpha = \frac{1}{2}$). The black thin dotted line marks the universal values for the minimum conductivity and the Fano factor.}}
\end{figure}

Finally, Tworzyd{\l}o \emph{et al.} show how conductivity and Fano factor evolute as a function of the width over length ratio using different boundary conditions \cite{tworzydlo2006}. In Figure \ref{evanescent-sigma-F} is reproduced the behavior of both conductivity and Fano factor as a function of $W/L$ for three different boundary conditions. At small $W/L$, both minimum conductivity and Fano factor take non-universal values. Note that calculations for zigzag edges have been done for the conductivity using tight binding theory \cite{robinson2007}, zigzag edges mixing different values of $k_{y,n}$ which strongly complicates the analytical solution.

\section{Experimental set-up, samples and shot noise measurement technique}

\begin{figure}[htbp]
\begin{center}\scalebox{0.5}{\includegraphics{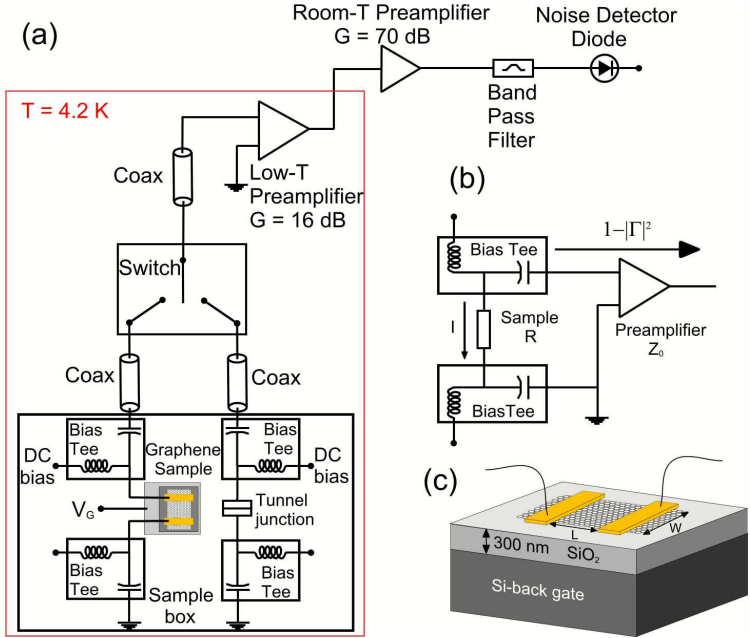}}\end{center}{\caption{\label{set-up}(a) Experimental set-up for detecting shot noise at $T$ = 4.2--30 K.
See the text for details. (b) Schematic of the principle of our measurements in terms of the
noise power reflection $|\Gamma|^2$. (c) Illustration of a typical
graphene sample fabricated for our noise study.}}
\end{figure}

Measuring shot noise requires carefully dedicated electronics. There are several ways to detect it such as cross correlation \cite{saminadayar1997,depicciotto1997,henny1999,oliver1999,roche2004,dicarlo2006} or SQUID-based resistance
bridge \cite{jehl2000} techniques. Depending on the nature of the studied system, one must avoid any low frequency noise known as $1/f$ noise (also called Flicker noise). By measuring the noise spectrum density at low frequency (10 Hz) $S_{I} = A\frac{\langle I^2 \rangle}{f^{\beta}}$ where $A$ is the the
noise amplitude coefficient and $\beta \sim 1$, we have extracted $A \sim 10^{-8}$ and checked that our set-up is well above the crossover frequency between $1/f$ and shot noise. In this work, we used a sensitive lock-in detection technique (see also \cite{wu2006,wu2007,danneau2008}), to improve the measurement sensitivity. The current is modulated using a sine-wave modulation, $I = I_{DC} + \delta I \sin(\omega t)$ where $I_{DC} \gg \delta I$, for the lock-in detection of noise. Alternatively, shot noise can also be detected using a dc set-up. In order to avoid external spurious signals, the set-up is placed in a Faraday cage. We use the shot noise generated by a tunnel junction, which is Poissonian ($\mathcal{F}$ = 1) to calibrate the graphene sample noise. The tunnel junctions are fabricated of Al/AlO$_{x}$/Al using standard two-angle shadow evaporation in an ultra-high vacuum system. A microwave switch is used to alternatively measure the noise from the graphene sample and the tunnel junction. We use bias-tees to split dc bias and the bias-dependent high-frequency noise signal. The noise signal is first amplified by a low-noise amplifier (LNA) with a noise temperature of $T_n$ = 3.5 K in matching conditions, thermalized at the same temperature as the sample. The noise detection scheme ends with a series of room-temperature amplifiers, and the signal is finally collected by a zero-bias Schottky diode with band-pass filtering of $f$ = 600--850 MHz to cut off EMI from mobile phone frequencies (see Figure \ref{set-up}(a)). All the data was measured in a helium dewar, in which samples were in a He-gas atmosphere of 1 bar. The resistance of the samples was measured using standard low-frequency ac lock-in technique with an excitation amplitude of 0.3 mV ($\sim$ 3 K) at $\frac{\omega}{2\pi} = 63.5$ Hz, in the temperature
range of 4.2--30 K.

\begin{table}[htbp]
\caption{Sample characteristics of the four samples measured in our experiments. $W/L$ is the width over length ratio. $\theta$ corresponds to the angle between the leads. $V_D$ defines the position of the Dirac point in gate voltage. These points were extrapolated from the minimum conductivity at $\frac{4e^2}{\pi h}$ for samples B and C. See text for more details.}
\label{tab:1}       % Give a unique label
\begin{center}\begin{tabular}{cccccc}
\hline\noalign{\smallskip}
Sample \verb"A" & Sample \verb"B" & Sample \verb"C"  & Sample \verb"D" & Sample \verb"E" & Sample \verb"F" \\
\noalign{\smallskip}\hline\noalign{\smallskip}
$\frac{W}{L}$ = 24 & $\frac{W}{L}$ = 10 & $\frac{W}{L}$ = 3 & $\frac{W}{L}$ = 2 & $\frac{W}{L}$ = 4.2 & $\frac{W}{L}$ = 1.8 \\
$L$ = 200 nm  & $L$ = 200 nm & $L$ = 300 nm & $L$ = 500 nm & $L$ = 950 nm & $L$ = 500 nm \\
$\theta$ = 0$^{\circ}$ & $\theta$ = 0$^{\circ}$ & $\theta$ = 0$^{\circ}$ & $\theta$ = 0$^{\circ}$ & $\theta$ = 0$^{\circ}$ & $\theta$ = 8$^{\circ}$ \\
$V_{D}$ = 19.5 V & $V_{D}$ = 145 V & $V_{D}$ = 100 V & $V_{D}$ = 78 V & $V_{D}$ = 28 V & $V_{D}$ = 22 V \\
\noalign{\smallskip}\hline
\end{tabular}
\end{center}
\end{table}

Graphene sheets are mechanically exfoliated using the Scotch tape technique and transferred from the graphite crystals (the graphite used here is a natural graphite powder) to the surface of a SiO$_{2}$/Si substrate (300 nm thick thermally grown SiO$_{2}$ layer). The silicon substrate is heavily doped and it is used as a back-gate (see Figure \ref{set-up}(c)). The single graphene layers are located using a three-CCD camera in an optical microscope on the base of the RGB green component shift \cite{oostinga2008}. After standard e-beam lithography, a bilayer Ti (10 nm) / Au (40 nm) is evaporated followed by lift-off with acetone. Chips are mounted in a homemade sample holder and micro-bonded with Al wire. We report measurements on four samples which are listed in Table \ref{tab:1}.

The noise power measured from the LNA is a mixture of thermal noise and the shot noise of the sample. It can be defined as a function of the reflected signal $|\Gamma|$. Here $|\Gamma| = \frac{|R-Z_{0}|}{|R+Z_{0}|}$ is the noise signal reflection coefficient when the noise source (the measured sample with a resistance $R$) does not match the circuit (here our cold amplifier is matched to a transmission line having an impedance of $Z_0$ = 50 $\mathrm{\Omega}$).
Then, the measured noise power can be expressed:
\begin{eqnarray}
P(I) &=& P_{noise}(1-|\Gamma|^{2}) = \mathcal{F} \times
2eV\frac{4RZ_0}{(R+Z_0)^{2}} \nonumber \\ &=& \mathcal{F} \times
2eI \times 4Z_0\left(\frac{R}{R+Z_0}\right)^{2},
\end{eqnarray}
where $P_{noise}=\mathcal{F}2eIR$ is the shot noise generated by the sample at $T=0$ (see Figure \ref{set-up}(b)).

In our experiments, we used a similar technique as in \cite{wu2006,wu2007,danneau2008} to measure the shot noise and extract the Fano factor. During the noise measurement the sample (with the differential resistance $R_{d} = \frac{dV}{dI}$) is coupled to the LNA with an impedance $Z_{0}$ = 50 $\mathrm{\Omega}$, where $i_{n}^{2}$ marks the full noise at the operating point, including the preamplifier noise and shot noise from the sample.

\begin{figure}[h]
\begin{center}\includegraphics[width=13pc]{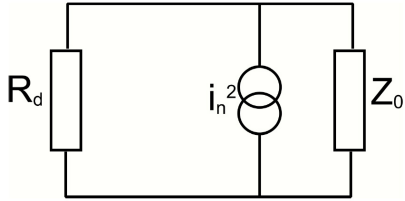}\hspace{2pc}%
\begin{minipage}[b]{9pc}\caption{\label{circuit}Schematic of the equivalent circuit of our measurement: $R_{d}$ and $Z_0$ represent the resistance of the sample and the cold preamplier respectively, $i_{n}^{2}$ represents the full noise generated by the circuit.}
\end{minipage}\end{center}
\end{figure}

We have used the electrical equivalent model shown in Figure \ref{circuit}, to calculate the coupling of the current fluctuations. Then, the noise power transferred to the cold amplifier can be written as:
\begin{eqnarray}
P_{Z_0}=\left(\frac{R_{d}}{R_{d}+Z_0}\right)^{2}Z_0i_{n}^{2}
\end{eqnarray}
and the measured noise signal is:
\begin{eqnarray}
P = gain \times BW \times g \times P_{Z_0} = \mathcal{G} \times P_{Z_0}\label{power}
\end{eqnarray}
where \emph{gain} refers to the total gain of the amplifier chain, \emph{BW} is the measurement bandwidth, \emph{g} denotes the sensitivity of the Schottky diode noise detector, and $\mathcal{G}$ is the calibration factor. In linear systems such as tunnel junctions or graphene, $R_{d}$ is constant.
Using our aforementioned set-up, we can write:
\begin{eqnarray}
\frac{1}{\mathcal{G}}\frac{1}{Z_0}\frac{\Delta P}{\Delta I} = \left(\frac{R_{d}}{R_{d}+Z_0}\right)^{2}\frac{\Delta i_{n}^{2}}{\Delta I}\label{first}
\end{eqnarray}
As the change of the noise generator $i_{n}^{2}$ is due to the shot noise part:
\begin{eqnarray}
\Delta i_{n}^{2} = \Delta S_{I} = 2eF_{d} \Delta I \label{second}
\end{eqnarray}
where $F_{d}$ is the differential Fano factor: then equation \ref{power} reads:
\begin{eqnarray}
\frac{1}{\mathcal{G}}\frac{1}{Z_0}\frac{\Delta P}{\Delta I} = \left(\frac{R_{d}}{R_{d}+Z_0}\right)^{2}2eF_{d}
\end{eqnarray}
Since usually $R_d \gg Z_0$, the coupling term $\frac{R_{d}}{R_{d}+Z_0}$ can be taken as 1.
From the tunnel junction measurement where $F = F_{d} = 1$ when  $eV \gg k_{B}T$, one can derive the calibration factor $\mathcal{G}$. After using it upon the graphene noise measurement, one gets the differential Fano factor $F_{d}$ as a function of the biasing current, and we then define the average Fano factor by integrating $F_{d}$ over the current bias:
\begin{eqnarray}
F = \frac{1}{I} \int_{0}^{I} F_{d}dI
\end{eqnarray}
which is the common Fano factor when $eV \gg k_{B}T$, and tends to zero around zero bias (due to thermal noise averaging). In terms of current noise, our average Fano factor corresponds to:
\begin{eqnarray}
F=\frac{S_I (I)-S_I (0)}{2eI}
\end{eqnarray}
For nonlinear system such as carbon nanotubes \cite{wu2007}, noise measurements are sensitive to changes in the sample resistance. We then have to take it into account and calculate corrections by deriving the differential resistance. If we now consider our sample with a differential resistance $R_d$ not constant, we must differentiate equation \ref{power}:
\begin{eqnarray}
\frac{1}{\mathcal{G}}\frac{1}{Z_0}\Delta P &=& \Delta \left[\left(\frac{R_{d}}{R_{d}+Z_0}\right)^{2}i_{n}^{2}\right]  \nonumber \\
&=& \left( \frac{R_{d}}{R_{d}+Z_0} \right)^{2} \Delta i_{n}^{2} + i_{n}^{2}\times 2\left(\frac{R_{d}}{R_{d}+Z_0}\right) \Delta \left(\frac{R_{d}}{R_{d}+Z_0}\right) \label{diff1}
\end{eqnarray}
%Then $\Delta (i_{n}^{2})$ can be written as equation \ref{second} with a first order correction:
%\begin{eqnarray}
%\Delta i_{n}^{2} = \left(\frac{R_{d}}{R_{d}+Z_0}\right)^{2}2eF_{d} \Delta I \sim \left(1-\frac{2Z_{0}}{R_{d}}\right)2eF_{d} \Delta I \label{diff2}
%\end{eqnarray}
and if we calculate:
\begin{eqnarray}
\Delta \left(\frac{R_{d}}{R_{d}+Z_0}\right) &=& \frac{(R_d+Z_0)\Delta R_d-R_d\Delta R_d}{(R_d+Z_0)^2}=\frac{Z_0}{(R_d+Z_0)^2}\Delta R_d   \label{diff3}
\end{eqnarray}
Then we can write equation \ref{diff1} as:
\begin{eqnarray}
\frac{1}{\mathcal{G}}\frac{1}{Z_0}\frac{\Delta P}{\Delta I} &=& \left( \frac{R_{d}}{R_{d}+Z_0} \right)^{2} \frac{\Delta (i_{n}^{2})}{\Delta I} + 2i_{n}^{2}\left(\frac{R_{d}}{R_{d}+Z_0}\right) \left(\frac{Z_0}{(R_d+Z_0)^2}\right)\frac{\Delta R_d}{\Delta I}
%\nonumber \\ &=& \frac{Z_0}{(R_d+Z_0)^2}\frac{\partial R}{\partial V}R_d = \frac{Z_0}{(R_d+Z_0)^2}\frac{\partial (\frac{\partial V}{\partial I})}{\partial V}R_d    \nonumber \\ &=& \frac{Z_0}{(R_d+Z_0)^2}\left(-R_d^2\frac{\partial ^2 I}{\partial V^2}\right)R_{d}
\label{diff2}
\end{eqnarray}
Since
\begin{eqnarray}
\frac{\Delta R_d}{\Delta I} &=& \frac{\partial R}{\partial V}R_d = \frac{\partial (\frac{\partial V}{\partial I})}{\partial V}R_d = \left(-R_d^2\frac{\partial ^2 I}{\partial V^2}\right)R_{d}
\label{diff3}
\end{eqnarray}
We finally obtain a new expression for equation \ref{diff1}:
\begin{eqnarray}
\frac{1}{\mathcal{G}}\frac{1}{Z_0}\frac{\Delta P}{\Delta I} \quad =  \quad 2eF_{d} \quad - \overbrace{2eF_{d}\frac{2Z_{0}}{R_{d}}}^{\mathrm{R_d}\ \mathrm{variation}} - \quad \overbrace{2i_n^2Z_0R_d\frac{\partial^{2}I}{\partial V^{2}}}^{\mathrm{total} \ \mathrm{system} \ \mathrm{noise}}
\end{eqnarray}
The first order correction comes from the measured shot noise due to $R_d$ variations and the second order corrections is caused by the total system noise due to the non-linearity, $i_n^2$ corresponding to the full noise at the operating point including the noise due to the LNA. Note that in our shot noise measurements in graphene, the correction is taken into account in the extraction of $F$ even though it is very small.

\begin{figure}[htbp]
\scalebox{0.24}{\includegraphics{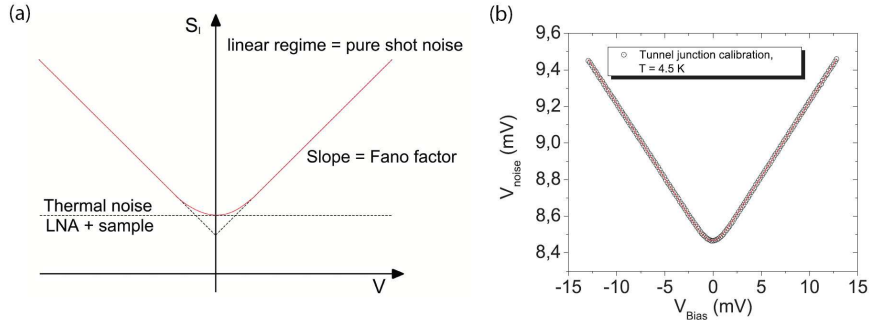}} {\caption{\label{curve-TJ} (a) Schematic of a typical noise curve showing the transition from thermal to pure shot noise. The noise given by the LNA shifts the curve to a higher noise level. (b) Typical tunnel junction noise measurement used for the Fano factor extraction of our graphene samples. The curve is fitted using the Khlus formula at $T$ = 4.5 K.}}
\end{figure}

Figure \ref{curve-TJ}(a) shows a schematic of a typical noise curve example that can be measured using our experimental set-up. As we mentioned in the previous parts of this article, shot noise gives information that cannot be extracted from classical dc transport measurements. However, it can be only detected if the frequency is high enough to overcome the $1/f$ noise. Shot noise occurs when the sample is biased; this is an out of equilibrium noise (also called excess noise). At low bias, thermal noise, originating from the random thermally excited vibration of the charge carriers, is predominant. Its spectral density can be defined as $S_I = kT\Delta f$, where $\Delta f$ is the noise bandwidth of the LNA. Thermal noise extends over all frequencies up to the quantum limit when $\hbar \omega > k_{B}T$ \cite{beenakker2003}. When the bias is large enough ($eV > k_BT$), the noise versus bias curve becomes linear and the detected noise is purely due to the shot noise. The slope of the linear part corresponds to the Fano factor. The minimum noise at zero bias is a mixture of the thermal noise of both the LNA and the sample.

Figure \ref{curve-TJ}(b) illustrates the very high resolution of our experimental set-up on a typical tunnel junction sample of resistance $R_{T}$ = 8 k$\mathrm{\Omega}$. The data are fitted using a formula similar to the one originally introduced by Khlus \cite{khlus1987} which describes the cross-over from thermal to shot noise when $eV \sim k_{B}T$ :
\begin{eqnarray}\label{equaKhlus}
S_{I}= \frac{4k_{B}T_{n}}{Z_{0}}+\mathcal{F}\frac{2e|V|R}{\left(R+Z_{0}\right)^2}
\coth\left(\frac{e|V|}{2k_{B}T} \right)\label{equa3}
\end{eqnarray}

where $T_n$ is the thermal noise of the LNA. Note that here, the thermal noise of the sample is neglected.

\section{Shot noise in graphene}

%The previous section dealt with the noise that occurs in conductors at low frequency.
Now we focus our work on shot noise. We used the experimental set-up and the technique to extract the Fano factor presented in Section 3. We have divided this section in three parts. In the first part, we will show that our measurements are well described by the evanescent wave theory and demonstrate that transport in graphene can be ballistic. In the second part, we will see how disorder affects the Fano factor and we will compare our findings with the existing theories modeling disordered graphene. Finally, we will show how non-parallel leads affect shot noise.

\subsection{Ballistic regime}

\begin{figure}[htbp]
\begin{center}\scalebox{0.5}{\includegraphics{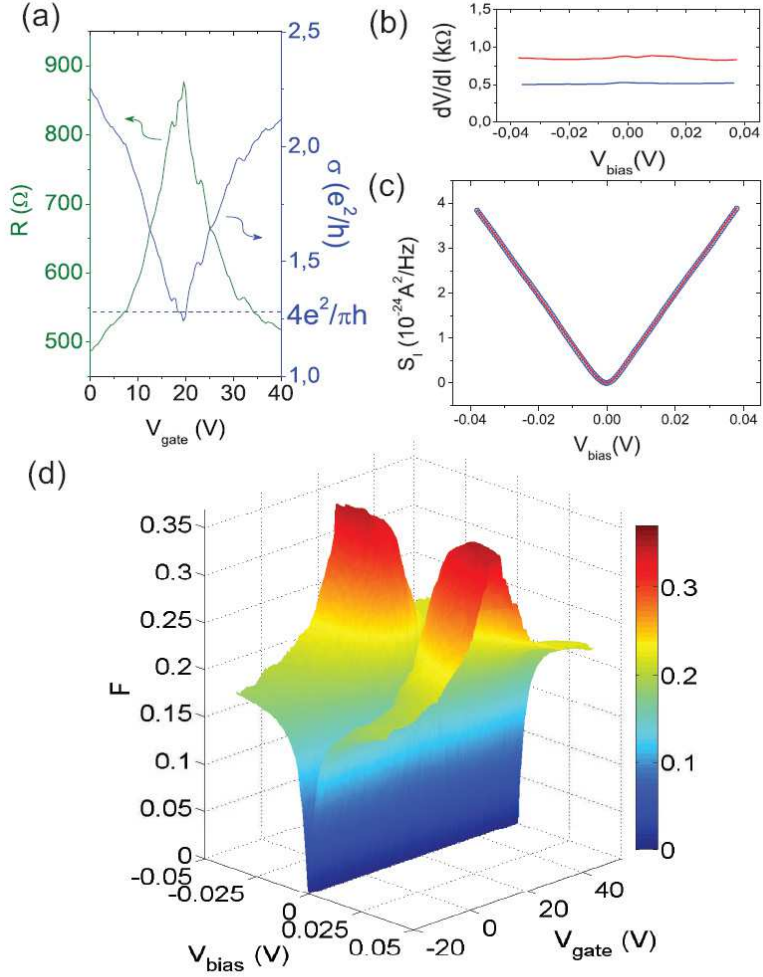}}\end{center}{\caption{\label{WoverL24}dc transport and shot noise measurements on
sample A: (a)  Resistance $R$ (left axis) and conductivity $\sigma$ (right axis) as a function of $V_{gate}$. (b) Differential resistance $dV/dI$ versus bias voltage $V_{bias}$ at the Dirac point (red curve) and at high density (blue curve). (c) Current noise per unit bandwidth $S_{I}$ as a function
of bias at the Dirac point, at $T$ = 8.5 K, fitted (red curve) using Khlus formula ($\mathcal{F}= $ 0.318). Note that the low-bias data are perfectly fitted as
well as the high-bias (d) Mapping of the average Fano factor $F$ as a function of gate voltage $V_{gate}$ and bias voltage $V_{bias}$ at $T =$
8.5 K.}}
\end{figure}

\begin{figure}[htbp]
\begin{center}\scalebox{0.55}{\includegraphics{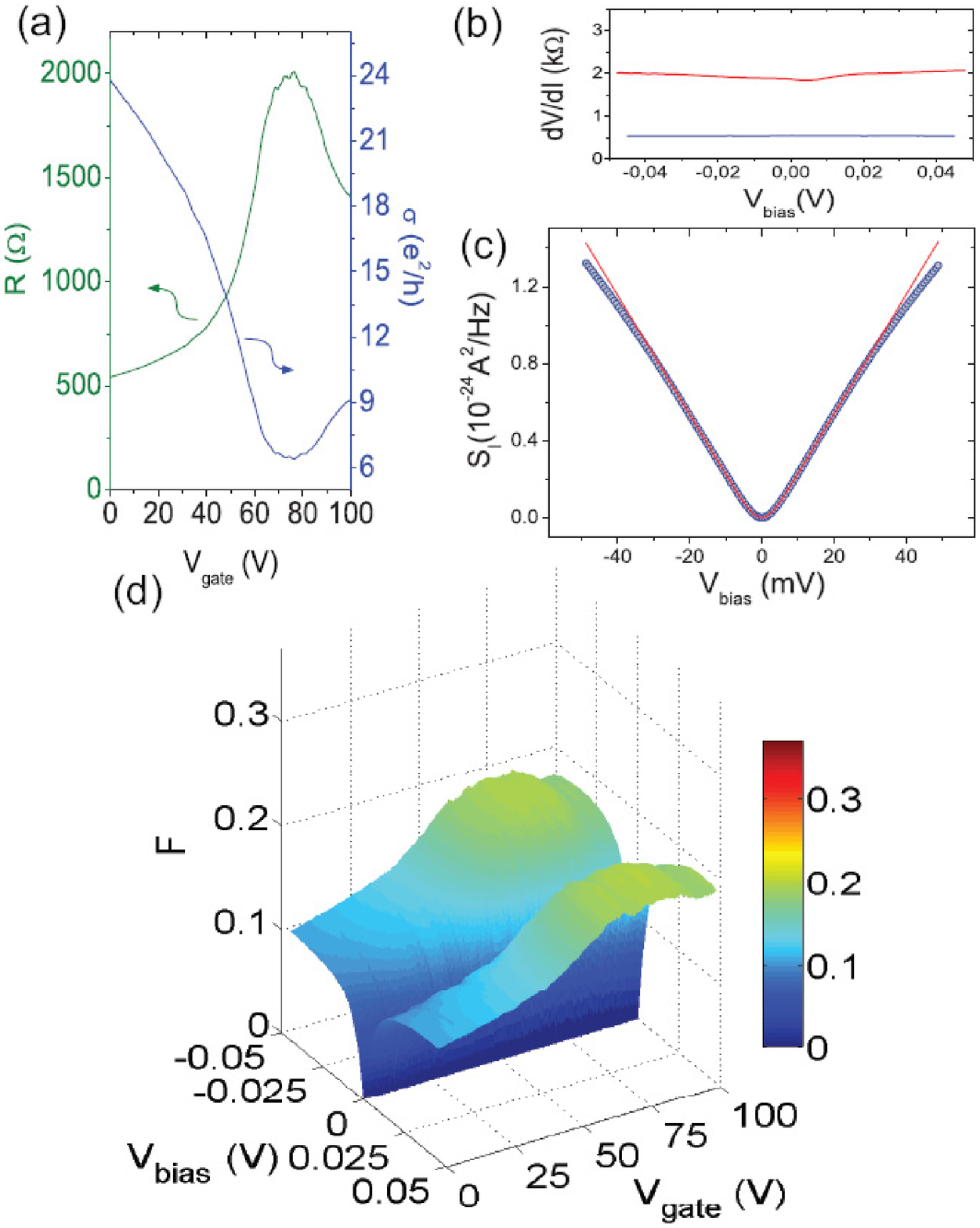}}\end{center}{\caption{\label{WoverL2}dc transport and shot noise measurements on
sample D : (a) Resistance $R$ (left axis) and conductivity $\sigma$ (right axis) as a function of gate voltage $V_{gate}$. (b) Differential resistance $dV/dI$ versus bias voltage $V_{bias}$ at the Dirac point (red curve) and at high density (blue curve). (c) Current noise per unit bandwidth $S_I$ as a function of the bias voltage $V_{bias}$ at the Dirac point, at $T =$ 5 K, fitted (red curve) using Khlus formula ($\mathcal{F}$ = 0.196). (d) Mapping of the average Fano factor $F$ as a function of the gate voltage $V_{gate}$ and the bias voltage $V_{bias}$ at $T =$ 5 K.}}
\end{figure}

We first present measurements on samples that have a distance between the leads $L \leq 500$ nm (i.e. samples \verb"A", \verb"B", \verb"C", \verb"D"). Sample \verb"A" has a very large aspect ratio. In Figure \ref{WoverL24}(a), we can see the resistance and conductivity of sample \verb"A" as a function of the gate voltage (i.e. charge carrier density). All of our graphene samples show a maximum resistance in positive gate voltage $V_{gate}$ values. This means that at zero gate voltage, the Fermi level lies in the valence band because our samples are non-intentionally \emph{p}-doped, probably due to oxygen gas adsorption \cite{schedin2007}. We clearly observe a maximum resistance and a minimum conductivity of around $\frac{4e^{2}}{\pi h}$ at the Dirac point, despite the chemical doping. From our measured conductivity values, it seems that adsorbed gas on a graphene sheet does not create strong scattering centers and thus does not affect dramatically the transport properties of our samples. For sample \verb"A", we obtain a minimum conductivity which is the one expected for large aspect ratio graphene strips \cite{tworzydlo2006} and observed experimentally in recent experiments \cite{miao2007}. It is important to note that the resistance of our graphene samples is nearly independent of the bias voltage $V_{bias}$, regardless of whether the measurement is taken at or far from the Dirac point (see Figure \ref{WoverL24}(b), as well as Figure \ref{WoverL2}(b), Figure \ref{WoverL4.2}(b) and Figure \ref{WoverL1.8}(b)). A non-linear behavior of the resistance, like in carbon nanotubes \cite{wu2007}, should be taken into account since noise measurements are sensitive to the sample resistance.

In Figure \ref{WoverL24}(c), we can see the current noise per unit bandwidth as a function of the $V_{bias}$ measured at the Dirac point at $T$ = 8.5 K in sample \verb"A". Using the Khlus formula (equation \ref{equaKhlus}) which describes the cross-over from thermal to shot noise when $eV \sim k_{B}T$, we have fitted and extracted the Fano factor $\mathcal{F}$ \cite{khlus1987}.

Since the resistance of our graphene samples is bias-independent, we may fit Khlus formula (equation \ref{equaKhlus}) to our data using only $\mathcal{F}$ as a fitting parameter at fixed temperature $T$. Note that when dealing with the integrated differential Fano factor, the excess noise $\frac{4k_{B}T_{n}}{Z_{0}}$ can be neglected. Using equation \ref{equaKhlus}, we have fitted and extracted the Fano factor $\mathcal{F} = 0.318$ at $T = 9$ K . We have also used our tunnel junction calibration technique to extract the average Fano factor $F$ \cite{wu2006,wu2007,danneau2008} to check the accuracy of our measurements. We found $F$ = 0.338 at the Dirac point (at $V_{bias}$ = 40 mV).

Our measurements seem to confirm that transport at the Dirac point may occur via evanescent waves \cite{katsnelson2006a,tworzydlo2006}. The two extracted Fano factors $\mathcal{F}$ and $F$ as well as the minimum conductivity are very close to the expected theoretical values of 1/3 and $\frac{4e^{2}}{\pi h}$ respectively at the Dirac point for a perfect graphene strip with large $W/L$ \cite{tworzydlo2006}. Note that the Fano factor in this case is also the one expected for a diffusive mesoscopic system at the Dirac point. In reference \cite{tworzydlo2006}, the authors demonstrate that the Fano factor should decrease as the charge carrier density increases which should not happen for a diffusive system (see the next subsection for the influence of disorder). In Figure  \ref{WoverL24}(d), we can see a mapping of the average Fano factor $F$, calculated by integrating the differential Fano factor $F_{d}$ as described in Section 3, as a function of the bias voltage $V_{bias}$ and the gate voltage $V_{gate}$. A clear dependence of $F$ on gate voltage (i.e. the charge carrier density) is observed, with a clear drop (about a factor of 2) of the Fano factor at large carrier density confirming, in turn, that our results are in good agreement with the evanescent state theory \cite{tworzydlo2006}, and that charge carriers in our sample do not undergo any inelastic scattering. Note that we cannot obtain a quantitative agreement with the evanescent mode theory, because doping by the leads may cause variation of the gate coupling capacitance and because the presence of non-uniform doping, that does not strongly scatter the charge carriers, also affects the electronic density of states. Nevertheless, the gate voltage scale is found to be larger than the one found in \cite{tworzydlo2006}. Comparing our data with a square lattice contact model in perfect graphene strips with large $W/L$ \cite{laakso2007}, we observe that the capacitance $C_{gate}$ in our sample is smaller by a factor of $\sim$ 9 compare to the one that gives a simple two infinite plane capacitor model (i.e. $C_{gate}$ $\sim$ 12 aF$\mu m^{-2}$ instead of 115 aF$\mu m^{-2}$).
We note that despite the presence of doping molecules on top of our graphene strip, the Fano factor remains equal to $\frac{1}{3}$, which is in a good agreement with a recent theory modeling fractal conductors \cite{groth2008}.

\begin{figure}[htbp]
\begin{center}\scalebox{0.25}{\includegraphics{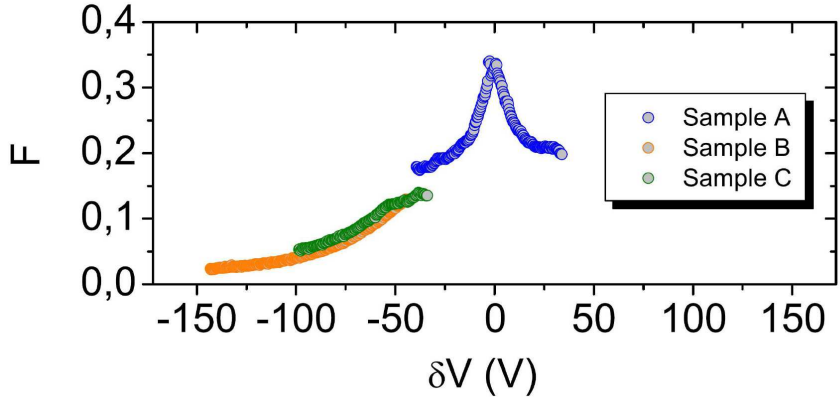}}\end{center}{\caption{\label{F-all}Average Fano factor $F$ extracted at $V_{bias}$ = 40 mV for samples A, B and C, all having $W/L \geq$ 3, as a function of $\delta V = V_{gate} - V_{Dirac}$, where $V_{Dirac}$ is the gate voltage value to reach the Dirac point. For the two unintentionally highly \emph{p}-doped samples (orange and green dots), the Dirac point was estimated via extrapolation of the minimum conductivity at $\frac{4e^2}{\pi h}$. At large $\delta V$, $F$ tends to 0, value expected for a ballistic mesoscopic system.}}
\end{figure}

We notice that the Fano factor is barely affected by temperature (up to $T$ = 30 K). This indicates that both the length $L$ and the width $W$ are smaller than the electron-phonon inelastic scattering length $L_{e-ph}$. If this condition were not fulfilled, the Fano factor would decrease, approximatively as inversely proportional to $N = \frac{L_{e-ph}}{ \max (W,L)}$ (note that the actual form of $L_{e-ph}$ would be model dependent \cite{blanter2000}). Since our shot noise measurements do not depend on temperature (between 4 and 30 K) and with our contacts being highly transparent, the presence of inelastic scattering mechanism in the graphene sample and at its interfaces with the leads can be ruled out. It is important to note that bad contacts can only increase the Fano factor toward the limit of two symmetrical tunneling barriers in series:
\begin{eqnarray}
\mathcal{F} = \frac{R_{T1}^{2} + R_{T2}^{2}}{(R_{T1}+R_{T2})^{2}}=\frac{1}{2}
\end{eqnarray}
This is not the case of our samples in which the Fano factor has never been measured higher than $\frac{1}{3}$.

In addition to sample \verb"A", we measured two other samples with large width over length ratios, samples \verb"B" and \verb"C" (all having $W/L\geq$ 3). The average Fano factor $F$ as a function of $\delta V = V_{gate} - V_{Dirac}$ is plotted in Figure \ref{F-all} for samples \verb"A", \verb"B" and \verb"C". All samples were \emph{p}-doped, the Dirac points being at positive gate voltages, but only for one of these three samples we could reach the Dirac point (sample \verb"A"). The gate voltages corresponding to the Dirac point for the two other samples were estimated from their conductivity curves. Despite the high doping level of the samples, the Fano factor seems to behave universally and tends to zero at very high density. This, indeed, demonstrates that graphene can behave as a ballistic conductor, contradicting recent calculations \cite{ziegler2007}. Despite the probable presence of some disorder in our system, the transport regime can be considered to be ballistic on our sample length scale.

We also measured sample \verb"D" which has a much smaller aspect ratio ($W/L =$ 2). In Figure \ref{WoverL2}(a), we can see that the minimum conductivity
reaches a much larger value than the sample with large $W/L$ ($\sim 6 \frac{e^{2}}{ h} \gg \frac{4e^{2}}{\pi h}$). We also verify that the resistance $R$ of the sample can be considered to be constant as a function of the bias (see Figure \ref{WoverL2}(b)). The spectral density of current noise as a function of $V_{bias}$ is shown in Figure \ref{WoverL2}(c). We observe that the data are well fitted at low bias using the Khlus formula. However, we can see a deviation at large bias which indicates here, a reduction of the Fano factor, presumably due to electron-phonon coupling \cite{blanter2000}. The Fano factor reaches $F =$ 0.196 at the Dirac point and eventually decreases substantially at large charge carrier density. A mapping of the average Fano factor $F$ as a function of bias $V_{bias}$ and gate voltage $V_{gate}$ is displayed in Figure \ref{WoverL2}(d). We observe that the determination of $\mathcal{F}$ in Figure \ref{WoverL2}(c) yields almost the same result: $\mathcal{F}= F = 0.19$ at the Dirac point. Our measurements are in good agreement with the results of \cite{tworzydlo2006} calculated for the case of metallic armchair edge for small $W/L$ for the Fano factor. Note that there is a discrepancy for the minimum conductivity which should be $\sim 1.1\frac{4e^2}{\pi h}$.

\subsection{Effect of disorder}

\begin{figure}[htbp]
\begin{center}\scalebox{0.5}{\includegraphics{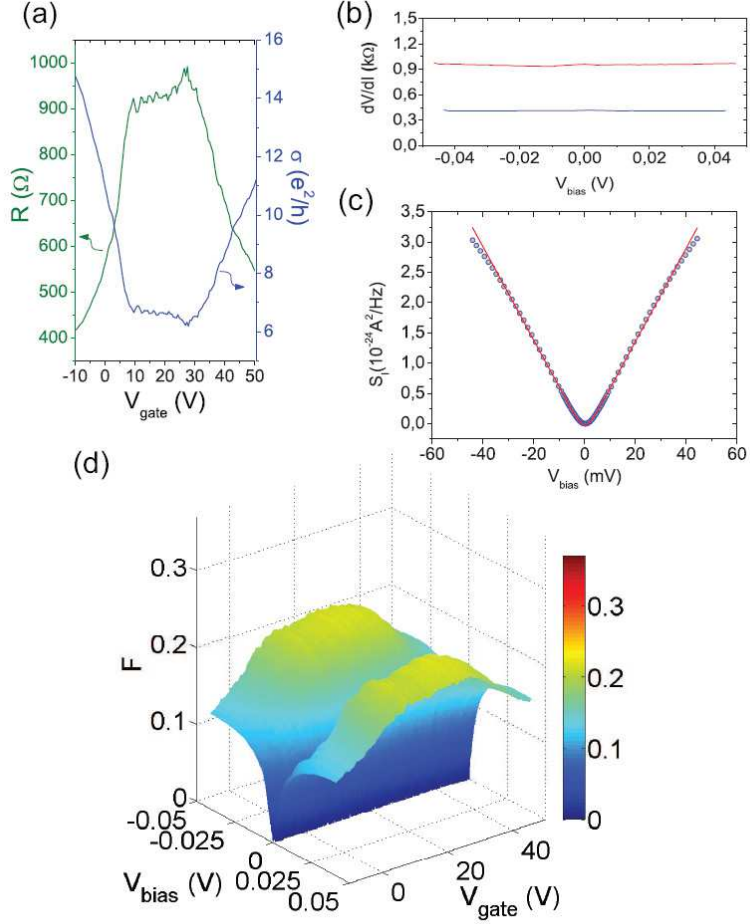}}\end{center}{\caption{\label{WoverL4.2}dc transport and shot noise measurements on
sample E: (a) Resistance $R$ (left axis) and conductivity $\sigma$ (right axis) as a function of gate voltage $V_{gate}$. (b) Differential resistance $dV/dI$ versus bias voltage $V_{bias}$ at the Dirac point (red curve) and at high density (blue curve). (c) Noise spectral density $S_I$ as a function of bias voltage
$V_{bias}$ at the Dirac point, at $T =$ 12 K, fitted (red curve) using Khlus formula ($\mathcal{F}$ = 0.256). (d) Mapping of the average Fano factor $F$ as a function of gate voltage $V_{gate}$ and bias voltage $V_{bias}$ at $T =$ 12 K.}}
\end{figure}

Disorder can dramatically influence electronic transport in mesoscopic conductors. Fano factor of $\frac{1}{3}$ has been predicted \cite{blanter2000} and measured in the case of diffusive systems \cite{henny1999a}. Multi-walled carbon nanotubes have shown some more exotic values, maybe due to electron-electron interactions in this one-dimensional system \cite{wu2007a}. Recent measurements in disordered graphene \cite{dicarlo2008} have shown a gate-independent Fano factor and a value higher than $\frac{1}{3}$ but smaller than $\frac{1}{2}$ (which is the maximum obtainable value for two symmetrical tunnel junctions), which could be due to bad graphene-contact interfaces (see previous section) or strong potential disorder \cite{lewenkopf2008}.

In the previous sections, we have seen that given the zero density of states at the Dirac point in graphene, transport occurs via evanescent modes instead of propagating modes \cite{tworzydlo2006}. Nevertheless, defects like vacancies \cite{peres2006} or dislocations \cite{carpio2008} can enhance locally the density of states at the Dirac point, which becomes finite, or create localized states which could create magnetic moments hampering charge carrier transport \cite{pereira2006}. As a consequence, in graphene with disorder, transport at the Dirac point might occur via a combination of evanescent and propagating modes due to presence of scattering. Indeed, perfect infinite two-dimensional graphene must be perfectly flat. Defects should curve this perfect plane and create roughness as in suspended membranes \cite{ledoussal1992}. Such ripples have been observed in suspended graphene \cite{meyer2007} and are believed to be intrinsic \cite{fasolino2007}. However, the origin of these corrugations on exfoliated graphene deposited on SiO$_{2}$/Si substrate \cite{ishigami2007,stolyarova2007} is still debated. Ripples have been clearly observed in graphene grown epitaxially on SiC  \cite{varchon2008}. In fact, it is believed that ripples could be the origin of the charge impurity formations at the Dirac peak \cite{dejuan2007,kim2007}. However, there is not yet a consensus as to the origin of these charge impurities \cite{dejuan2007,kim2007,hwang2007,nomura2006,morozov2008}. Such charge puddles \cite{martin2007} creating potential scattering centers should influence conduction in graphene \cite{hwang2007}. Another limiting factor in the carrier mobility is the interaction with the substrate \cite{fratini2008}. It was shown that intrinsic carrier mobility could be as high as 200 000 cm$^2$V$^{-1}$s$^{-1}$ \cite{morozov2008}. This was achieved using suspended structure \cite{bolotin2008} after current annealing \cite{moser2007}. However, mobility achievable in graphene on SiO$_{2}$/Si substrate is on the order of $\mu$ = 20 000 cm$^2$V$^{-1}$s$^{-1}$ so far, corresponding to elastic mean free path $l_e \sim$ 500 nm at a carrier density $n = 5 \times 10^{12}$ cm$^{-2}$ \cite{tan2007}.

Recent theories showed that disorder should enhance conductivity in graphene via impurity resonant tunneling \cite{titov2007}. Such counterintuitive behavior can be understood as a consequence of the absence of intervalley scattering \cite{morpurgo2006} and the chirality conservation \cite{katsnelson2006b}. It was also shown that weak disorder may induce anomalously large conductance fluctuations at high charge carrier density \cite{rycerz2007}. By modeling smooth potential disorder, San Jose \emph{et al.} have shown that near the Dirac point at length scales $\ll L,W$, disorder increases the minimum conductivity and lowers the Fano factor at the Dirac point, down to 0.243 for one-dimensional disorder and to 0.295 for the two-dimensional case \cite{sanjose2007}. A diffusive system should not display any gate dependence. This was demonstrated for long-range disorder in \cite{lewenkopf2008}. A gate-dependent Fano factor appears once the disorder strength is reduced.

We have measured shot noise in sample \verb"E" which has a large $W/L$ and large distance between the leads, approaching 1 $\mu$m. In Figure \ref{WoverL4.2}(a), the resistance $R$ and the conductivity $\sigma$ are plotted versus gate voltage $V_{gate}$. The resistance curve is not as peaked as it should be, and in fact, the Dirac point seems to be truncated probably due to the presence of disorder. Note that the graphene sheet is, again, \emph{p}-doped. We also see that the minimum conductivity is no longer $\frac{4e^{2}}{\pi h}$ but much larger, which is also in agreement with the fact that disorder should increase the conductivity  in graphene \cite{titov2007}. In Figure \ref{WoverL4.2}(b), we see that the resistance remains constant when the bias is tuned.

From our noise measurements, we observe a strong decrease of the Fano factor at the Dirac point compared to $\frac{1}{3}$ expected by the evanescent wave theory. Figure \ref{WoverL4.2}(c) shows the noise spectral density measured on sample \verb"E" at $T$ = 12 K. Using Khlus formula, we extract a Fano factor at the Dirac point $\mathcal{F}$ = 0.256. The fit is not perfect at high bias probably due to electron-phonon coupling. We also note that the curve is slightly asymmetrical. The average Fano factor gives a smaller value of about 0.23 (see Figure \ref{WoverL4.2}(d)). These values are in good agreement with the model which takes into account one dimensional smooth potential disorder \cite{sanjose2007}. In Figure \ref{WoverL4.2}(d), we observe that the Fano factor is reduced by tuning the gate voltage $V_{gate}$, proving that the disorder present in our sample in smoother than in \cite{dicarlo2008}.

\subsection{Non-parallel leads}

\begin{figure}[htbp]
\begin{center}\scalebox{0.55}{\includegraphics{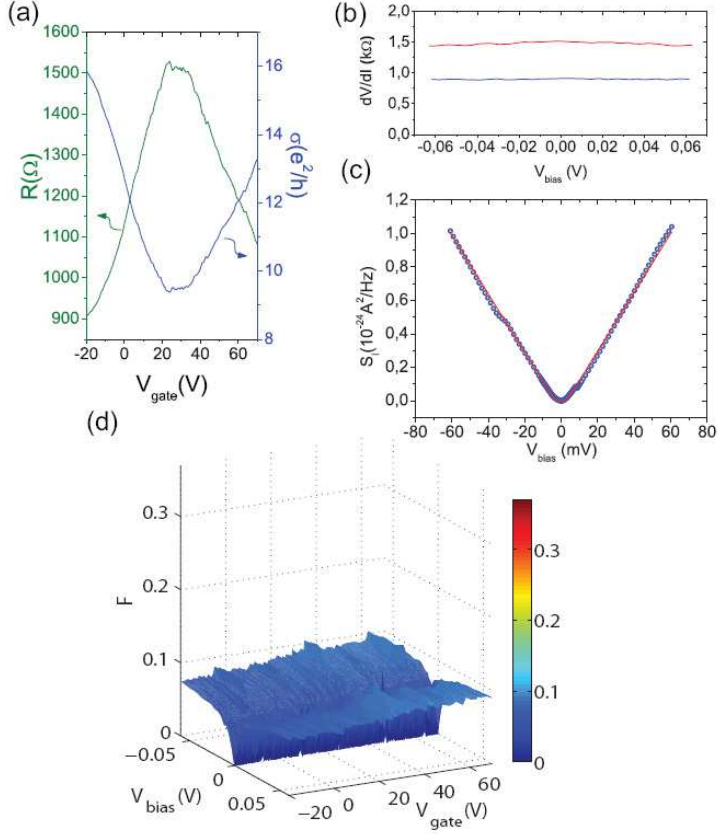}}\end{center}{\caption{\label{WoverL1.8}dc transport and shot noise measurements on
sample F: (a) Resistance $R$ (left axis) and conductivity $\sigma$ (right axis) as a function of gate voltage $V_{gate}$. (b) Differential resistance $dV/dI$ versus bias voltage $V_{bias}$ at the Dirac point (red curve) and at high density (blue curve). (c) Noise spectral density $S_I$ as a function of bias voltage
$V_{bias}$ at the Dirac point, at $T =$ 12 K, fitted (red curve) using Khlus formula ($\mathcal{F}$ = 0.087). (d) Mapping of the average Fano factor $F$ as a function of gate voltage $V_{gate}$ and bias voltage $V_{bias}$ at $T =$ 12 K.}}
\end{figure}

In this last part, we present measurements on conductivity and shot noise in sample \verb"F" which has non-parallel leads. In the frame of the evanescent wave theory for metallic armchair edges and rectangular samples, one would expect to measure a sub-Poissonian shot noise smaller than $\frac{1}{3}$ (around $\mathcal{F} \sim 0.17$ for  $W/L$ = 1.8). Figure \ref{WoverL1.8}(a) displays the resistance $R$ and the conductivity $\sigma$ as a function of gate voltage $V_{gate}$, showing a maximum resistance at a positive gate voltage, sample \verb"F" being \emph{p}-doped. We note that the minimum conductivity is much higher than the value predicted by the evanescent wave theory. Figure \ref{WoverL1.8}(b) confirms that the graphene strip behaves as an ohmic system. In Figure \ref{WoverL1.8}(c), we can see current noise per unit bandwidth $S_{I}$ as a function of bias voltage $V_{bias}$ at the Dirac point. We note that the noise seems to be a bit unstable. Using Khlus formula, we have fitted the data and extracted a Fano factor at the Dirac point $\mathcal{F}$ = 0.087. This value has been confirmed by extracting the average Fano factor $F$. Surprisingly, by tuning the gate voltage  $V_{gate}$ we do not observe that the Fano factor is maximum at the Dirac point and varies when the charge carrier density is tuned. As shown in Figure \ref{WoverL1.8}(d), the Fano factor is barely affected by the gate voltage which is similar to what has been as seen by DiCarlo \emph{et al.} \cite{dicarlo2008}, but with a much larger Fano factor value $> 0.35$.

In the model describing transport occurring via evanescent waves at the Dirac point \cite{katsnelson2006a,tworzydlo2006}, the condition for quantization of the transverse wave vector is essential to finally obtain the transmission eigenvalue distribution. The lead cross-section determines the transverse part of the wave function. If the leads are non-parallel or not rectangular, one can expect that the condition for quantization of the transverse and longitudinal modes are no longer fulfilled. The mixing of the transverse and the longitudinal modes may change the distribution function of the transmission probabilities of the evanescent states, acting similarly as the effect of disorder \cite{robinson2007}. Consequently, both conductivity and Fano factor should be modified by non-parallel leads. However, to our knowledge, there is no model available considering the effect of sample geometry on conductivity and Fano factor.

\section{Conclusions}

We have studied transport and noise in graphene strips. Shot noise measurements were performed in four different cases. We have seen a gate-dependent shot noise in short graphene strips with large and small $W/L$. At the Dirac point, we observed that for large $W/L$ both minimum conductivity and Fano factor reach universal
values of $\frac{4e^2}{\pi h}$ and 1/3 respectively. At very large density, the Fano factor tends to zero which is the value expected for a ballistic system. For $W/L$ smaller than 3, the Fano factor is lowered and the minimum conductivity increases. These findings are well explained by the evanescent wave theory describing transport at the Dirac point in perfect graphene. When $L$ is large enough, we see a significant reduction of the Fano factor at the Dirac point, reaching a value of 0.23, which is in good agreement with recent models taking into account smooth potential disorder like charge puddles \cite{sanjose2007}. Finally, in the case of non-parallel contacts, quantized transverse modes being mixed up, the measured Fano factor is no longer maximum at the Dirac point and almost gate-voltage independent, even though a clear maximum resistance is observed.

\begin{acknowledgements}
We thank A. Castro Neto, Y. Hancock, A. Harju, T. Heikkil\"a, A. K\"arkk\"ainen, M. Laakso, C. Lewenkopf, E. Mucciolo, M. Paalanen, P. Pasanen, E. Sonin, P. Virtanen and J. Wengler for fruitful discussions.
\end{acknowledgements}


\begin{thebibliography}{}

\bibitem{geim2007} A.K. Geim, and K.S. Novoselov, Nature Mat. \textbf{6}, 183 (2007).

\bibitem{katsnelson2006a} M.I. Katsnelson, Eur. Phys. J. B \textbf{51}, 157 (2006).

\bibitem{tworzydlo2006} J. Tworzyd{\l}o, B. Trauzettel, M. Titov, A. Rycerz, and C.W.J. Beenakker, Phys. Rev. Lett. \textbf{96}, 246802 (2006).

\bibitem{blanter2000} Ya. M. Blanter, and  M. B\"uttiker, Phys. Rep. \textbf{336}, 1 (2000).

\bibitem{saminadayar1997} L. Saminadayar, D. C. Glattli, Y. Jin and B. Etienne, Phys. Rev. Lett. \textbf{79}, 2526 (1997).

\bibitem{depicciotto1997} R. de Picciotto, M. Reznikov, M. Heiblum, V. Umansky, G. Bunin, and D. Mahalu, Nature \textbf{389}, 162 (1997).

\bibitem{henny1999} M. Henny, S. Oberholzer, C. Strunk, T. Heinzel, K. Ensslin, M. Holland, and C. Sch\"onenberger, Science \textbf{284}, 296 (1999).

\bibitem{oliver1999} W.D. Oliver, J. Kim, R.C. Liu, Y. Yamamoto, Science \textbf{284}, 299 (1999).

\bibitem{safonov2003} S.S. Safonov, A.K. Savchenko, D.A. Bagrets, O.N. Jouravlev, Y.V. Nazarov, E.H. Linfield, and D.A. Ritchie, Phys. Rev. Lett. \textbf{91}, 136801 (2003).

\bibitem{roche2004} P. Roche, J. S\'{e}gala, D.C. Glattli, J.T. Nicholls, M. Pepper, A.C. Graham, K.J. Thomas, M.Y. Simmons, and D.A. Ritchie, Phys. Rev. Lett. \textbf{93}, 116602 (2004).

\bibitem{dicarlo2006} L. DiCarlo, Y. Zhang, D. T. McClure, D.J. Reilly, C.M. Marcus, L.N. Pfeiffer, and K.W. West, Phys. Rev. Lett. \textbf{97}, 036810 (2006).

\bibitem{dattabook} S. Datta, Electronic transport in mesoscopic systems, Cambridge University Press (1995).

\bibitem{beenakker1992} C.W.J. Beenakker and M. B\"uttiker, Phys. Rev. B \textbf{46}, 1889 (1992).

\bibitem{nagaev1992} K. Nagaev, Phys. Lett. A \textbf{169}, 103 (1992).

\bibitem{shimuzu1992} A. Shimizu and M. Ueda, Phys. Rev. Lett. \textbf{69}, 1403 (1992).

\bibitem{steinbach1996} A.H. Steinbach, J.M. Martinis, and M.H. Devoret, Phys. Rev. Lett. \textbf{76}, 3806 (1996).

\bibitem{schomerus2007} H. Schomerus,  Phys. Rev. B, \textbf{76} 045433 (2007).

\bibitem{sonin2008} E.B. Sonin, Phys. Rev. B \textbf{77}, 233408 (2008).

\bibitem{novoselov2005} K.S. Novoselov, A.K. Geim, S.V. Morozov, D. Jiang, M.I. Katsnelson, I.V. Grigorieva, S.V. Dubonos, and A.A. Firsov, Nature \textbf{438}, 197 (2005).

\bibitem{tan2007} Y.-W. Tan, Y. Zhang, K. Bolotin, Y. Zhao, S. Adam, E.H. Whang, S. Das Sarma, H.L. Stormer, and P. Kim, Phys. Rev. Lett. \textbf{99}, 246803 (2007).

\bibitem{badarson2007} J.H. Bardarson, J. Tworzyd{\l}o, P.W. Brouwer, and C.W.J. Beenakker, Phys. Rev. Lett. \textbf{99}, 106801 (2007).

\bibitem{laakso2008} M.A. Laakso and T.T. Heikil\"{a}, arXiv:08064528.

\bibitem{snyman2007} I. Snyman and C.W.J. Beenakker, Phys. Rev. B, \textbf{75} 045322 (2007).

\bibitem{katsnelson2006} M.I. Katsnelson, Eur. Phys. J. B \textbf{52}, 151 (2006).

\bibitem{cheianov2006} V.V. Cheianov and V.I. Fal'ko, Phys. Rev. B \textbf{74}, 041403(R) (2006).

\bibitem{abanin2007} D.A. Abanin and L.S. Levitov, Science \textbf{317}, 641 (2007).

\bibitem{robinson2007} J.P. Robinson and H. Schomerus, Phys. Rev. B \textbf{76}, 115430 (2007).

\bibitem{jehl2000} X. Jehl, M. Sanquer, R. Calemczuk, and D. Mailly, Nature \textbf{405}, 50 (2000).

%\bibitem{heersche2007} H.B. Heersche, P. Jarillo-Herrero, J.B. Oostinga, L.M.K. Vandersypen, A.F. Morpurgo, Nature \textbf{446}, 56 (2007).

\bibitem{wu2006} F. Wu, L. Roschier, T. Tsuneta, M. Palaanen, T.H. Wang, and P.J. Hakonen, AIP Conf. Proc. \textbf{850}, 1482 (2006).

\bibitem{wu2007} F. Wu, P. Queipo, A. Nasibulin, T. Tsuneta, T.H. Wang, E. Kauppinen and P.J. Hakonen, Phys. Rev. Lett. \textbf{99}, 156803 (2007).

\bibitem{danneau2008} R. Danneau, F. Wu, M.F. Craciun, S. Russo, M.Y. Tomi, J. Salmilehto, A.F. Morpurgo, and P.J. Hakonen, Phys. Rev. Lett. \textbf{100}, 196802 (2008).

\bibitem{oostinga2008} J.B. Oostinga, H.B. Heersche, X. Liu, A.F. Morpurgo, and L.M.K. Vandersypen, Nature Mat. \textbf{7}, 151 (2008).

\bibitem{beenakker2003} C.W.J. Beenakker and C. Sch\"onenberger, Phys. Today \textbf{56}, 37 (2003).

\bibitem{khlus1987} V.A. Khlus, Zh. Ekps. Teor. Fiz. \textbf{93}, 2179 (1987) [Sov. Phys. JETP \textbf{66}, 1243 (1987)].

%\bibitem{motchenbacherbook} C.D. Motchenbacher and J.A. Connelly,  Low-noise electronic system design, Wiley, (1993).

%\bibitem{hooge1969} F.N. Hooge, Phys. Lett. \textbf{29A}, 139 (1969).

%\bibitem{collins2000} P.G. Collins, M.S. Fuhrer, and A. Zettl, Appl. Phys. Lett. \textbf{76}, 894 (2000).

%\bibitem{ishigami2006} M. Ishigami, J.H. Chen, E.D. Williams, D. Tobias, Y.F. Chen, and M.S. Fuhrer, Appl. Phys. Lett. \textbf{88}, 203116 (2006).

%\bibitem{collins2000a} P.G. Collins, K. Bradley, M. Ishigami, and A. Zettl, Science \textbf{287}, 1801 (2000).

\bibitem{schedin2007} F. Schedin, A.K. Geim, S.V. Morozov, E.W. Hill, P. Blake, M.I. Katsnelson, and K.S. Novoselov, Nature Mat. \textbf{6}, 652 (2007).

%\bibitem{liu2006} F. Liu, K.L. Wang, D. Zhang, and C. Zhou, Appl.Phys. Lett. \textbf{89}, 063116 (2006).

%\bibitem{tobias2008} D. Tobias, M. Ishigami, A. Tselev, P. Barbara, E.D. Williams, C.J. Lobb, and M.S. Fuhrer, Phys. Rev. B \textbf{77}, 033407 (2008).

%\bibitem{zchen2007} Z. Chen, Y.-M. Lin, M.J. Rooks, and P. Avouris, Physica E \textbf{40}, 228 (2007).

%\bibitem{lin2008} Y.-M. Lin, and P. Avouris, Nanolett. (in press).

\bibitem{miao2007} F. Miao, S. Wijeratne, Y. Zhang, U.C. Coskun, W. Bao, and C.N. Lau, Science \textbf{317}, 1530 (2007).

\bibitem{laakso2007} M.A. Laakso, private communication.

\bibitem{groth2008} C.W. Groth, J. Tworzyd{\l}o and C.W.J. Beenakker, Phys. Rev. Lett. \textbf{100}, 176804 (2008).

\bibitem{ziegler2007} K. Ziegler, Physica E \textbf{40}, 2622 (2007).

\bibitem{henny1999a} M. Henny, S. Oberholzer, C. Strunk, and C. Sch\"{o}nenberger, Phys. Rev. B \textbf{59}, 2871 (1999).

\bibitem{wu2007a} F. Wu, T. Tsuneta, R. Tarkiainen, D. Gunnarsson, T.-H. Wang, and P.J. Hakonen, Phys. Rev. B \textbf{75}, 125419 (2007).

\bibitem{dicarlo2008} L. Dicarlo, J.R. Williams, Y. Zhang, D.T. McClure, and C.M. Marcus, Phys. Rev. Lett. \textbf{100}, 156801 (2008).

\bibitem{lewenkopf2008} C.H. Lewenkopf, E.R. Mucciolo, and A.H. Castro Neto, Phys. Rev. B \textbf{77}, 081410(R) (2008).

\bibitem{peres2006} N.M.R. Peres, F. Guinea, and A.H. Castro Neto, Phys. Rev. B \textbf{73}, 125411 (2006).

\bibitem{carpio2008} A. Carpio, L.L. Bonilla, F. de Juan, and M.A.H. Vozmediano, New J. Phys. \textbf{10}, 053021 (2008).

\bibitem{pereira2006} V.M. Pereira, F. Guinea, J.M.B. Lopes dos Santos, N.M.R. Peres, and A.H. Castro Neto, Phys. Rev. Lett. \textbf{96}, 036801 (2006).

\bibitem{ledoussal1992} P. Le Doussal and L. Radzihovsky, Phys. Rev. Lett. \textbf{69}, 1209 (1992).

\bibitem{meyer2007} J.C. Meyer, A.K. Geim, M.I. Katsnelson, K.S. Novoselov, T.J. Booth, and S. Roth, Nature \textbf{446}, 60 (2007).

\bibitem{fasolino2007} A. Fasolino, A.H. Los, and M.I. Katsnelson, Nature Mat. \textbf{6}, 858 (2007).

\bibitem{ishigami2007}  M. Ishigami, J.H. Chen, W.G. Cullen, M.S. Fuhrer, and E.D. Williams, Nano Lett. \textbf{7}, 1643 (2007).

\bibitem{stolyarova2007} E. Stolyarova, K.T. Rim, S. Ryu, J. Maultzsch, P. Kim, L.E. Brus, T.F. Heinz, M.S. Hybertsen, and G.W. Flynn, Proc. Natl Acad. Sci. USA \textbf{104}, 9209–9212 (2007).

\bibitem{varchon2008} F. Varchon, P. Mallet, J.-Y. Veuillen, and L. Magaud, Phys. Rev. B \textbf{77}, 235412 (2008).

\bibitem{dejuan2007} F. de Juan, A. Cortijo, and M.A.H. Vozmediano, Phys. Rev. B \textbf{76}, 165409 (2007).

\bibitem{kim2007} E.-A. Kim, and A.H. Castro Neto, arXiv:0702562.

\bibitem{hwang2007} E.H. Hwang, S. Adam, and S. Das Sarma, Phys. Rev. Lett. \textbf{98}, 186806 (2007).

\bibitem{nomura2006} K. Nomura, A.H. MacDonald, Phys. Rev. Lett. \textbf{96}, 256602 (2006).

\bibitem{morozov2008} S.V. Morozov, K.S. Novoselov, M.I. Katsnelson, F. Schedin, D.C. Elias, J.A. Jaszczak, and A.K. Geim, Phys. Rev. Lett. \textbf{100}, 016602 (2008).

\bibitem{martin2007} J. Martin, N. Akerman, G. Ulbricht, T. Lohmann, J.H. Smet, K. von Klitzing, and A. Yacoby, Nature Phys. \textbf{4}, 144 (2008).

\bibitem{fratini2008} S. Fratini and F. Guinea, Phys. Rev. B, \textbf{77} 195415 (2008).

\bibitem{bolotin2008} K.I. Bolotin, K.J. Sikes, Z. Jiang, M. Klima, G. Fudenberg, J. Hone, P. Kim, H.L. Stormer, Solid State Commun. \textbf{146}, 351 (2008).

\bibitem{moser2007} J. Moser, A. Barreiro, and A. Bachtold, Appl. Phys. Lett. \textbf{91}, 163513 (2007).

\bibitem{titov2007} M. Titov, Europhys. Lett. \textbf{79}, 17004 (2007).

\bibitem{morpurgo2006} A.F. Morpurgo, and F. Guinea, Phys. Rev. Lett. \textbf{97}, 196804 (2006).

\bibitem{katsnelson2006b} M.I. Katsnelson, K.S. Novoselov, and A.K. Geim, Nature Phys. \textbf{2}, 620 (2006).

\bibitem{rycerz2007} A. Rycerz, J. Tworzyd{\l}o, and C.W.J. Beenakker, Europhys. Lett. \textbf{79}, 57003 (2007).

\bibitem{sanjose2007} P. San-Jose, E. Prada and D.S. Golubev, Phys. Rev. B, \textbf{76} 195445 (2007).

\end{thebibliography}


\end{document}